\documentclass{webofc}
\usepackage[varg]{txfonts}   
\begin{document}
\title{Molecular dynamic simulation of water vapor interaction with blind pore of dead-end and saccate type}
%
%

\author{\firstname{Eduard G.} \lastname{Nikonov}\inst{1}\fnsep\thanks{\email{e.nikonov@jinr.ru}} \and
        \firstname{Miron} \lastname{Pavlu\v{s}}\inst{2}\fnsep\thanks{\email{miron.pavlus@unipo.sk}} \and
        \firstname{M\'aria} \lastname{Popovi\v{c}ov\'a}\inst{2}\fnsep\thanks{\email{maria.popovicova@unipo.sk}}
}

\institute{Joint Institute for Nuclear Research,\ 141980\ Dubna,\ Moscow Region,\ Russia 
\and
           University of Pre\v{s}ov,\ str. Kon\v{s}tant\'inova 16,\ 080 01 Pre\v{s}ov,\ Slovakia 
          }

\abstract{%
One of the varieties of pores, often found in natural or artificial building materials, are the so-called blind pores of dead-end or saccate type. Three-dimensional model of such kind of pore has been developed in this work. This model has been used for simulation of water vapor interaction with individual pore by molecular dynamics in combination with the diffusion equation method. Special investigations have been done to find dependencies between thermostats implementations and conservation of thermodynamic and statistical values of water vapor - pore system. The two types of evolution of water – pore system have been investigated: drying and wetting of the pore. Full research of diffusion coefficient, diffusion velocity and other diffusion parameters has been made.
}
\maketitle
\section{Molecular dynamics model}
\label{sec-1}
In classical molecular dynamics, the behavior of an individual particle is described by the Newton equations of motion \cite{Gould}.
For a simulation of particle interaction we use the Lennard-Jones potential \cite{LJ} with $\sigma = 3.17 \mbox{\AA}$ and $\varepsilon = 6.74\cdot 10^{-3} eV$. It is the most used to describe the evolution of water in liquid and saturated vapor form. Equations of motion 
were integrated by Velocity Verlet method \cite{Verlet}.
Berendsen thermostat \cite{Berendsen} is used for temperature equilibrating and control. Coefficient of the velocity recalculation $\lambda(t)$ at every time step $t$ depends on the so called ‘rise time’ of the thermostat $\tau_B$ which belongs to the interval $ [0.5,2] \ \mbox{ps}$. 
The Berendsen algorithm is simple to implement and it is very efficient for reaching the desired temperature from far-from-equilibrium configurations.
\section{Computer simulation of microscopic model}
\label{sec-2}
We made simulation for a pore of dimensions  $l_x=500$ nm,\ $l_y=50$ nm,\ $l_z=50$ nm 
with integration time step $\Delta t=0.016$ ps and evolution time 65.3 ns. Otherwise, we have considered the following input data for the drying process: 
\begin{itemize}
\item
1000 $H_2O$ molecules in the pore volume $500\times 50\times 50$ nm$^3$ form
a saturated water vapor at temperature $T_0=25$ $^oC$ and pressure $p_0=3.17$ $kPa$;
\item
1800 molecules in the outer area that form 20 \% of the saturated water vapor 
\end{itemize}
and input data for the wetting process:
\begin{itemize}
\item
200 $H_2O$ molecules in the pore volume $500\times 50\times 50$ nm$^3$ are 20 \%  of saturated water vapor;
\item
9000 molecules in the outer area form
a saturated water vapor at temperature $T_0=25$ $^oC$ and pressure $p_0=3.17$ $kPa$.
\end{itemize}

\begin{figure}[ht]
\center{\includegraphics[width=0.48\linewidth]{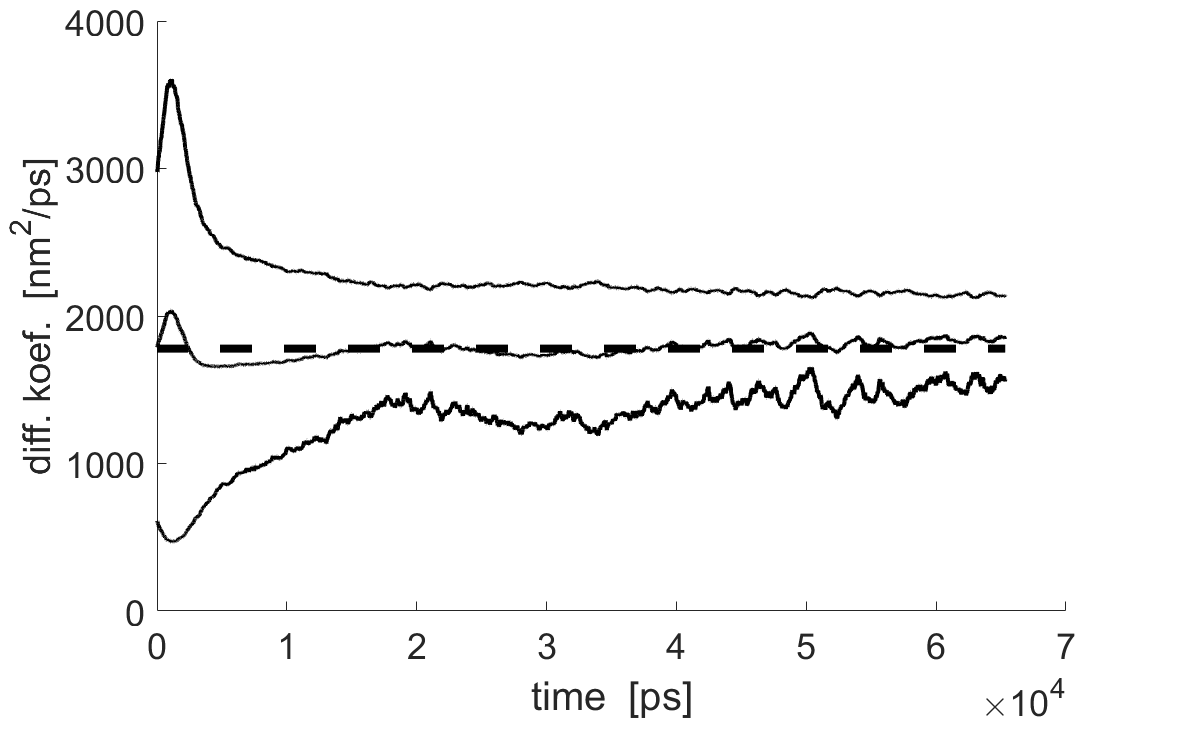}
\includegraphics[width=0.48\linewidth]{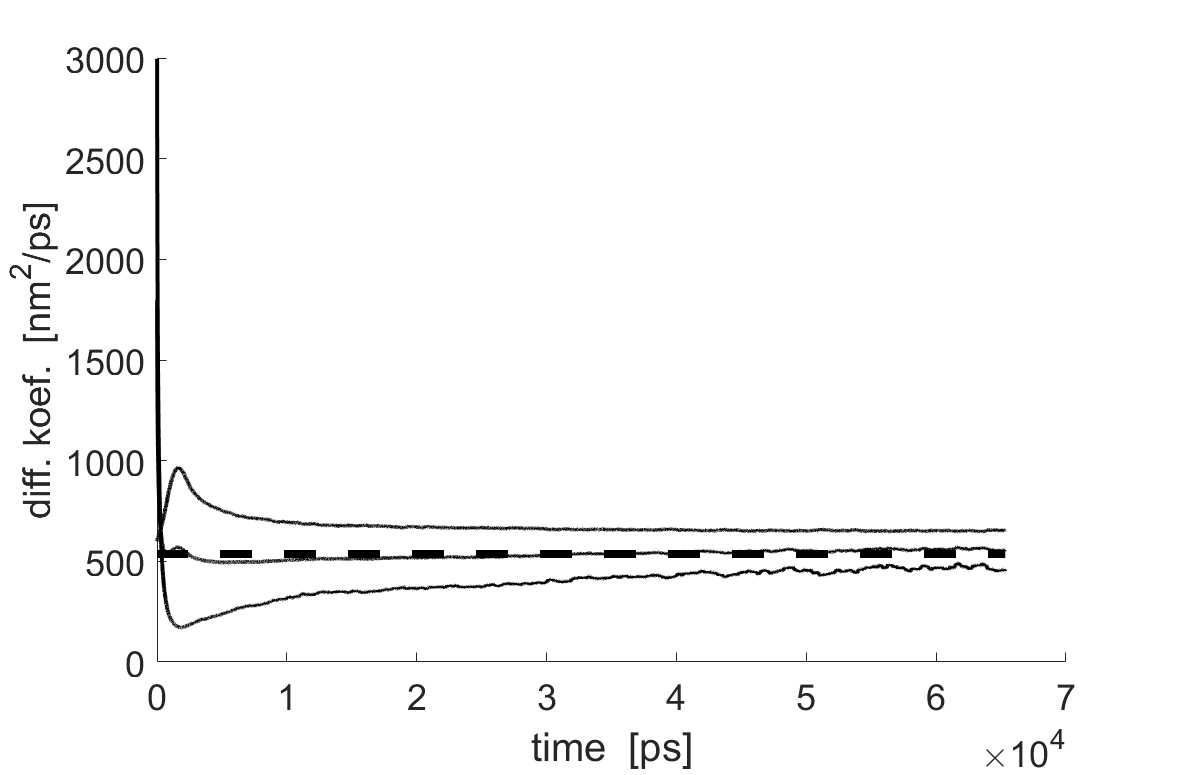}}
\caption{Diffusion coefficients for drying (left) and for wetting (right) processes (upper curves in pore, down curves in outer space, middle curves - the mean of pore and outer space, and dashed lines - constant values)
}
\label{Diffusion_Coef}
\end{figure}

The diffusion coefficients for drying process (left) and for wetting process (right) are shown in Fig. \ref{Diffusion_Coef}. The left figure depicts diffusion coefficients for pore (upper curve), for outer area (down curve), for mean value of previous (middle curve) and for constant value $D=1779.1$ [nm$^2$/ps] (middle dashed line). The right figure shows diffusion coefficients for pore (upper curve), for outer area (down curve), for mean value of previous (middle curve) and for constant value $D=536.33$ [nm$^2$/ps] (middle dashed line).
 
\section{Macroscopic diffusion model}
\label{sec-3}
Let us denote the water vapor concentration as $w_v(x,y,z,t)$ [$ ng/(nm)^3$] where $x,y,z$ are space independent variables and $t$ is time independent variable. Then, we consider the following macroscopic diffusion model
\begin{equation}\label{eq01}
\frac{\partial w_v}{\partial t}=D\Big(\frac{\partial^2 w_v}{\partial x^2}+\frac{\partial^2 w_v}{\partial y^2}+\frac{\partial^2 w_v}{\partial z^2}\Big)
\end{equation}
$$
0<x<l_x,\quad 0<y<l_y,\quad 0<z<l_z,\quad t>0
$$ 
\begin{equation}\label{eq02}
w_v(x,y,z,0)=w_{v,0},\qquad 0\leq x\leq l_x,\quad 0\leq y\leq l_y,\quad 0\leq z\leq l_z
\end{equation} 
\begin{equation}\label{eq03}
\left.\frac{\partial w_v}{\partial n}(t)\right|_{(x,y,z)\in\Gamma_2\cup\Gamma_3\cup\Gamma_4\cup\Gamma_5\cup\Gamma_6}=0,\qquad t>0
\end{equation} 
\begin{equation}\label{eq04}
\left.-D\frac{\partial w_v}{\partial x}(t)\right|_{(l_x,y,z)\in\Gamma_1}=\beta [w_v(l_x,y,z,t)-w_{v,out}(t)]
\end{equation} 
$$
0\leq y\leq l_y,\quad 0\leq z\leq l_z,\quad t>0
$$
where $D$ is the diffusion coefficient [$(nm)^2/ps$]; $l_x, l_y, l_z$ are 3D pore dimensions [$nm$]; $\Gamma_1,\Gamma_2,\Gamma_3,\Gamma_4,\Gamma_5,\Gamma_6$ are boundaries of 3D pore ($\Gamma_1$ is free boundary while the rest boundaries are isolated; $w_{v,0}$ is the initial concentration of water vapor, $w_{v,0}=2.304\cdot 10^{-17}$  for the drying process, and $w_{v,0}=0.461\cdot 10^{-17}$ for the wetting process [$ng/(nm)^3$]; $w_{v,out}(t)$ is the water vapor concentration in outer area [$ng/(nm)^3$]; $\beta$ is the coefficient of water vapor transfer from pore space to outer space, $\beta=50000$ [$nm/ps$]. 

We suppose that the outer area water vapor concentration is expressed as
$
w_{v,out}(t)=\varphi_0\cdot w_{sv}(T_0), 
$       
where $\varphi_0$ is the relative humidity of outer space ($0\le\varphi_0\le 1$) and $w_{sv}(T_0)$ is saturated water vapor concentration at outer temperature $T_0$.  

The linear problem (\ref{eq01})--(\ref{eq04}) can be solved exactly using the variables separation method \cite{Bitsadze} 
and the result of the solution is  
\begin{eqnarray}\label{eq05}
w_v(x,y,z,t)=w_{sv}(T_0)\cdot\varphi_0+\Big[w_{v,0}-w_{sv}(T_0)\cdot\varphi_0\Big]\cdot \\ \cdot\sum_{m=1}^{\infty}\sum_{n=0}^{\infty}\sum_{p=0}^{\infty}e^{\lambda_{mnp} D t}c_{mnp}\cos(\alpha_{xm} x)\cos(\alpha_{yn} y)\cos(\alpha_{zp} z) \nonumber
\end{eqnarray}
$$
\qquad 0\leq x\leq l_x,\qquad 0\leq y\leq l_y,\qquad 0\leq z\leq l_z,\qquad t>0.
$$ 
Here, $c_{mnp}$ are coefficients of unity expansion
$$
c_{mnp}=
  \begin{cases}
    \frac{4\sin(\alpha_{xm} l_x)}{2 l_x\alpha_{xm}+\sin(2\alpha_{xm} l_x)}       & \quad \text{if } n=0;\ p=0;\ m=1,2,3,\dots\\
    0                                                                                                                          & \quad \text{if } m,n,p=1,2,3,\dots\\
  \end{cases}
$$
and $\lambda_{mnp}$ are eigenvalues where 
$$
\lambda_{mnp}=-\alpha_{xm}^2-\alpha_{yn}^2-\alpha_{zp}^2,  
$$
$$
\alpha_{yn}=\frac{n\pi}{l_y},\qquad n=0,1,2,\dots\qquad\qquad\qquad\alpha_{zp}=\frac{p\pi}{l_z},\qquad p=0,1,2,\dots
$$
and $\alpha_{xm}$ are solutions of the equation
$$
\alpha_{xm}\cdot\tan(\alpha_{xm} l_x)=\beta/D,\qquad m=1,2,3,\dots. 
$$

\begin{figure}[ht]
\center{\includegraphics[width=0.48\linewidth]{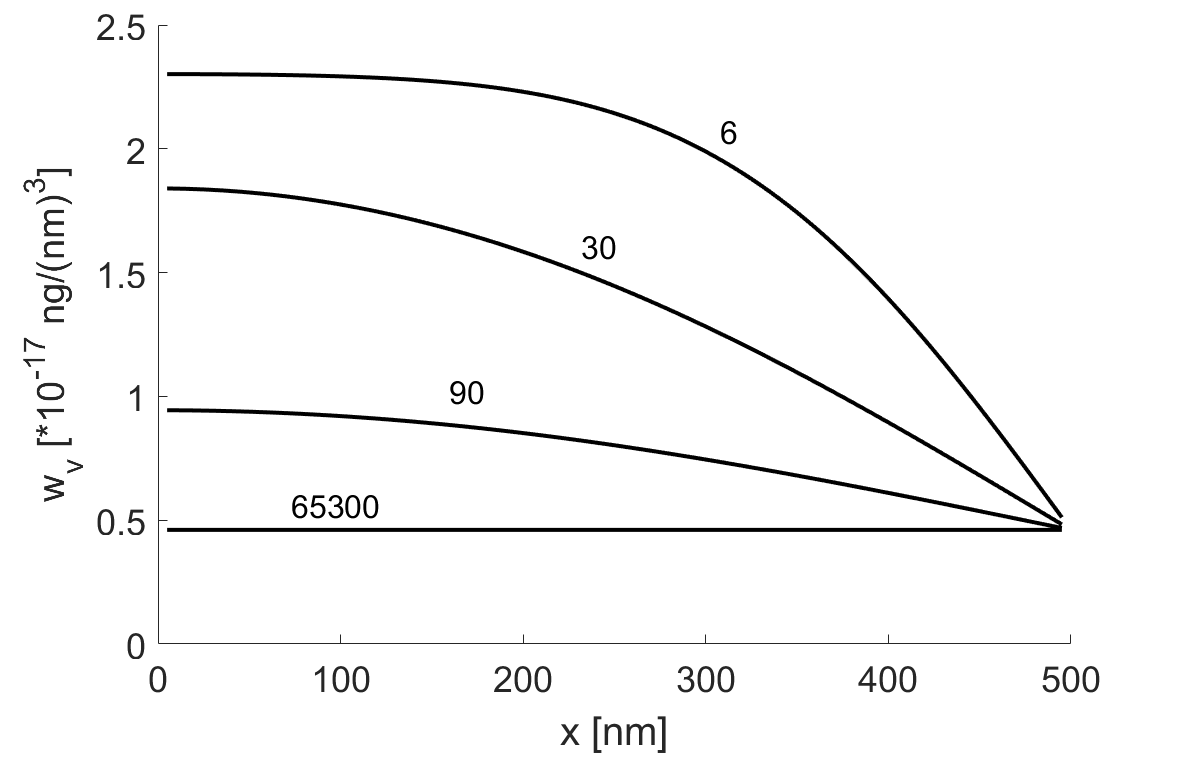}\includegraphics[width=0.48\linewidth]{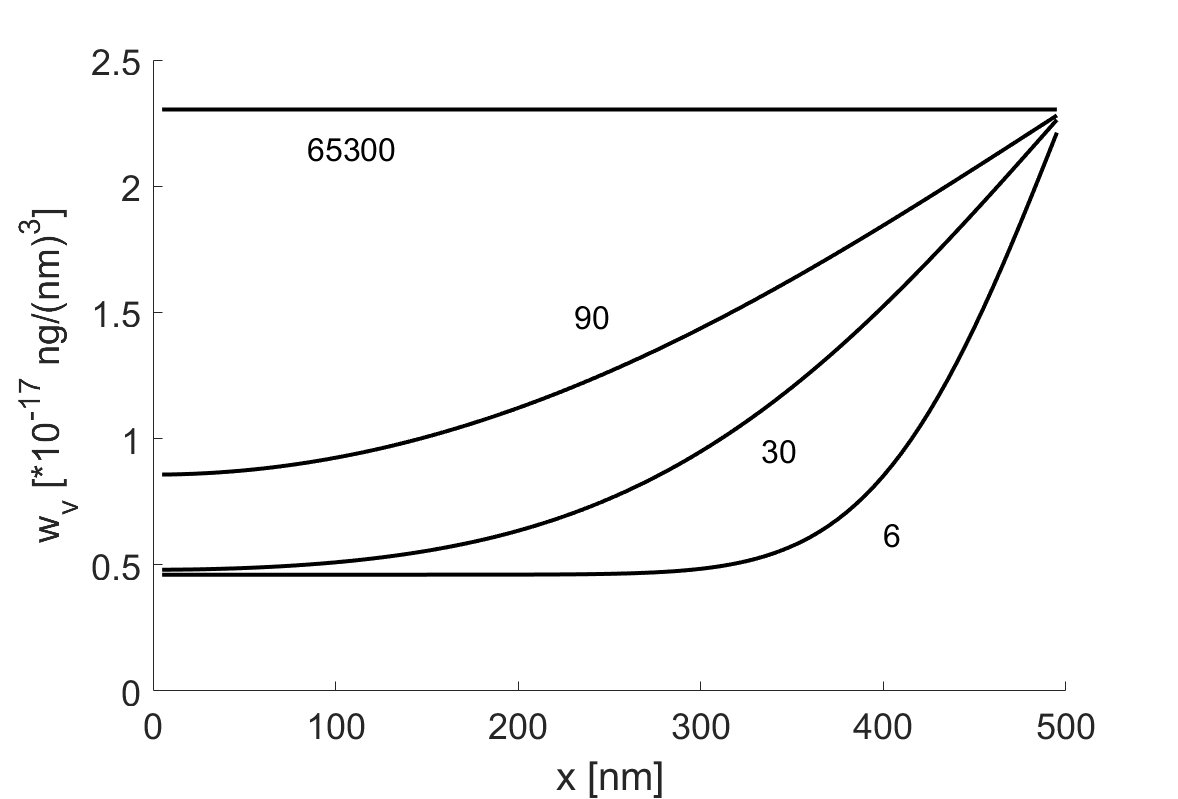}}
\caption{The dynamics of water vapor concentration at different time moments $t=6,\ 30,\ 90,\ 65300$ ps (cross section at $y=l_y/2$ and $z=l_z/2$), drying process (left) from top to bottom, and wetting process (right) from bottom to top}
\label{Dynamics_wv_ly_deleno_2_lz_deleno_2}
\end{figure} 
Results of macro model calculations are presented in Fig. \ref{Dynamics_wv_ly_deleno_2_lz_deleno_2}.
\section{Comparison of microscopic and macroscopic models}
\label{sec-4}
\begin{figure}[ht]
\center{\includegraphics[width=0.48\linewidth]{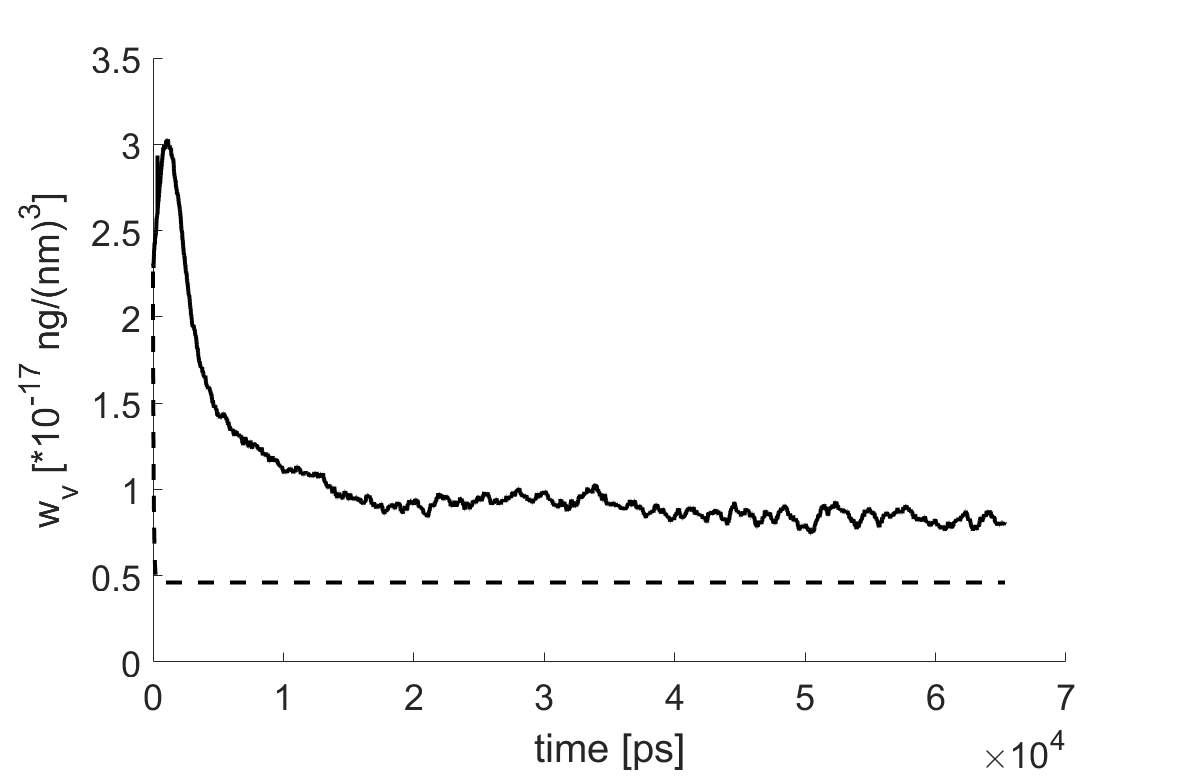}
\includegraphics[width=0.48\linewidth]{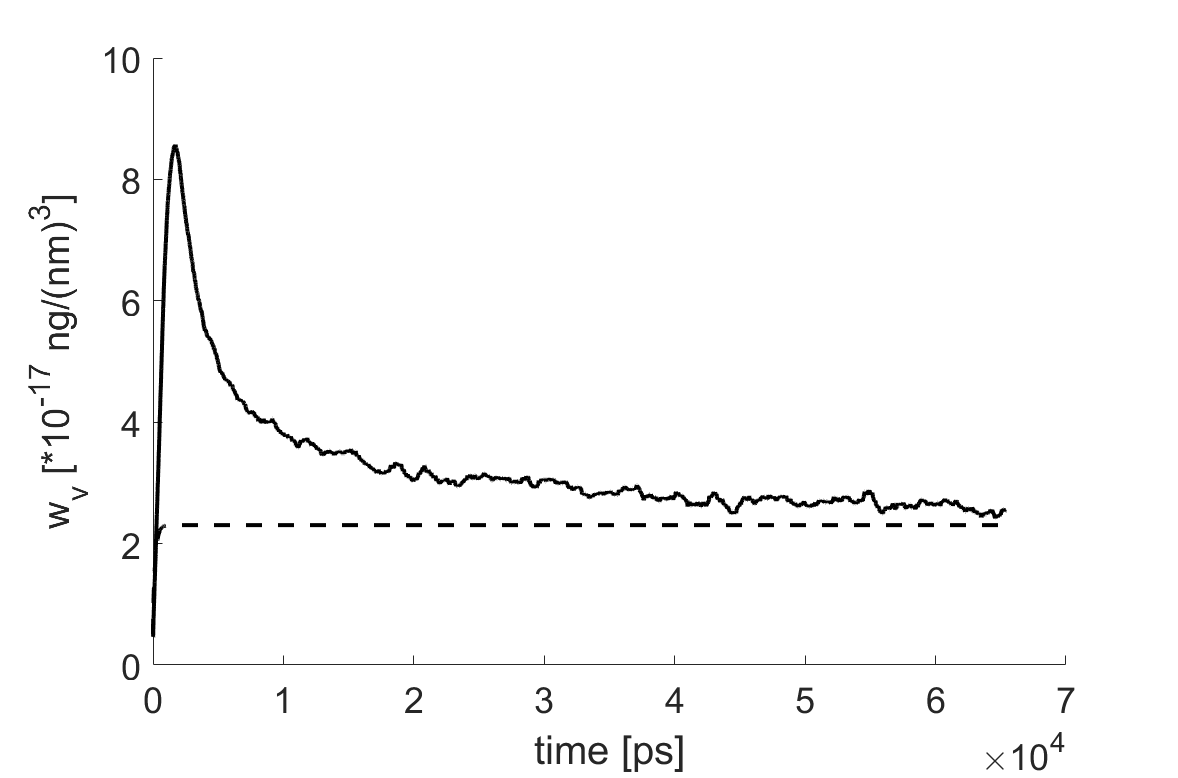}}
\caption{The dynamics of water vapor concentration (WVC) obtained by diffusion equation (dashed curves) versus WVC obtained by molecular dynamics (solid curves). The drying process (left) and the wetting process (right).}
\label{wv-rho}
\end{figure}

Finally, we compare the space mean value 
$$
\frac{1}{l_xl_yl_z}\int_{0}^{l_x}\int_{0}^{l_y}\int_{0}^{l_z}w_v(x,y,z,t)\ dxdydz
$$
of water vapor concentration (\ref{eq05}) for macro model with the density obtained by micro model. The results are in the Fig.\ref{wv-rho}.   

Our investigations allow to affirm that an approach based on combination of diffusion coefficients determination by means of molecular dynamics and further application of  these coefficients in macro model computations is useful for accuracy increasing of water-pore interaction description.

\end{document}